\documentclass[12pt,aps,prb,preprint]{revtex4} 
\usepackage[latin1]{inputenc}
\usepackage{amsmath,amssymb,amsfonts,amsthm}
\usepackage{verbatim}
\usepackage{epsf}
\usepackage{epsfig}
\DeclareGraphicsExtensions{.eps}
\usepackage{color}
\usepackage{makeidx}  

\newtheorem{definition}{Definition}
\newtheorem{proposition}{Proposition}

\begin{document}

\title{On the QKD relaying models}
\author{Quoc-Cuong Le 
and Patrick Bellot 
}
\affiliation{
Institut TELECOM, Telecom ParisTech, Paris, France\\
37-39 rue Dareau, 75014 Paris, France
} 
\date{\today}

\begin{abstract}
  We investigate Quantum Key Distribution (QKD) relaying models.
  Firstly, we propose a novel quasi-trusted QKD relaying model. The
  quasi-trusted relays are defined as follows: (i) being honest enough
  to correctly follow a given multi-party finite-time communication
  protocol; (ii) however, being under the monitoring of eavesdroppers.
  We develop a simple 3-party quasi-trusted model, called Quantum
  Quasi-Trusted Bridge (QQTB) model, to show that we could securely
  extend up to two times the limited range of single-photon based QKD
  schemes. We also develop the Quantum Quasi-Trusted Relay (QQTR)
  model to show that we could securely distribute QKD keys over
  arbitrarily long distances. The QQTR model requires EPR pair sources,
  but does not use entanglement swapping and entanglement purification
  schemes as proposed in \cite{WDHJB99,HKLHFC08,DCNG05}. Secondly, we
  show that our quasi-trusted models could be improved to become
  untrusted models in which the security is not compromised even
  though attackers have full controls
  over some relaying nodes. We call our two improved models the
  Quantum Untrusted Bridge (QUB) and Quantum Untrusted Relay
  (QUR) ones. The QUB model works on single photons and allows
  securely extend up to two times the limited QKD range. The QUR model
  works on entangled photons but does not use entanglement swapping
  and entanglement purification operations. This model allows securely transmit
  shared keys over arbitrarily long distances without dramatically
  decreasing the key rate of the original QKD schemes.
\end{abstract}

\maketitle

\section{Introduction}

The limited range of Quantum Key Distribution (QKD) link is one of
the most headache-questions to many researchers for a long time.
The earliest QKD protocol \cite{BB84} is the BB84 protocol that had
been proposed by Bennett and Brassard in 1984. After, this protocol was
proven to be unconditionally secure
\cite{PSJP00,HKL01,Chau02,MAYER01}, and promised the vast potentially
worthful applications. As the cost due of its extremely good security,
unfortunately, QKD owns undesirable restrictions over range and
rate \cite{QKDrate,SPI150}. This explains why there are few practical QKD
applications so far. Today, improving QKD range's approaches can be
roughly divided into two categories. The first one is improvements over direct QKD links, for instance, perfecting quantum
devices as quantum sources and quantum detectors. The second one is QKD relaying methods that allow to securely relay QKD keys. This
paper addresses the latter one. We will assume that we work with perfect
quantum devices, free-error quantum channels to focus
on the ``relaying'' aspect.

Our main contributions are :
\begin{enumerate}
\item The proposal of a new concept called ``quasi-trusted relay''
  that seems reasonable in realistic scenarios,
\item The Quantum Quasi-Trusted Bridge (QQTB) model that allows to
  securely extend up to two times the QKD range without invoking
  entanglement-based operations,
\item The Quantum Quasi-Trusted Relay (QQTR) model that allows to
  securely distribute shared keys over arbitrarily long distances
  without invoking entanglement swapping and entanglement purification
  operations,
\item The Quantum Untrusted Bridge (QUB) model that allows to securely
  extend up to two times the QKD range without invoking
  entanglement-based operations,
\item The Quantum Untrusted Relay (QUR) model that allows to securely
  distribute shared keys over arbitrarily long distances without
  invoking entanglement swapping and entanglement purification as
  proposed in \cite{WDHJB99,HKLHFC08,DCNG05}.
\end{enumerate}

The remainder is organized as follows. Section \ref{sec:rw_motiv}
gives an overview of previous works on QKD relaying models and
introduces our motivation. Section \ref{sec:bck_grd} reminds some
background concept and also makes some propositions that are used in 
our proposed models. We define our ``quantum
quasi-trusted'' concept in Section \ref{sec:qqtr_def}. Section
\ref{sec:qqtb_model} develops the Quantum Quasi-Trusted Bridge (QQTB)
model that is capable of
securely doubling the range of single-photon based QKD schemes. 
Section \ref{sec:qqtr_model} develops the Quantum-Trusted Relay (QQTR)
model that is capable of securely distributing shared keys
over arbitrarily long distances. Section
\ref{sec:qub_model} develops the Quantum Untrusted Bridge (QUB)
model that is capable of
securely doubling the range of single-photon based QKD schemes, in
releasing all constraints of the relaying node.
Section \ref{sec:qur_model} develops the Quantum Untrusted Relay (QUR)
model that is capable of securely distributing shared keys
over arbitrarily long distances, in releasing all constraints of
relaying nodes. We conclude in
Section~\ref{sec:conc}.

\section{\label{sec:rw_motiv}Related work and motivation}
\subsubsection{Related work.}
Since the range of QKD is limited, QKD relaying methods are
necessary. This becomes indispensable when one
aims at building QKD networks as in the last recent years. All QKD relaying methods so far
introduce some undesirable drawbacks. The most practical
method is based on trusted model. This method has been applied in two famous
QKD networks, DAPRA and SECOCQ \cite{BBN1,BBN2,RAFR06,RA06}. In this method, all the relaying nodes must
be assumed perfectly secured. Such an assumption is critical since passive
attacks or eavesdropping on intermediate nodes are very difficult to
detect. A few number of intermediate nodes could lead to a great
vulnerability in practice. Consequently, one wants to limit the
number of trusted nodes in QKD networks.

One could claim that the idea of the quasi-trusted QKD relaying model
is not new. The works in \cite{RIVF07,NTMS07,ISPEC08} were indeed
based on such an idea. However, the ``quasi-trusted'' property was
characterized differently and had been analyzed in a different
context: each node was assumed to be trusted with a high probability
$p\sim 1$, and the main focus was the security behavior of the global
system that consists of a great number of nodes. In this paper, we
propose a very different concept of ``quasi-trusted'' that is
characterized by: (i) being honest enough to correctly follow a given
multi-party finite-time communication protocol; (ii) however, being
under the monitoring of eavesdroppers.
 
Theoretically, the most strong QKD relaying model so far is the one
that is based on entanglement swapping (QS) operation
\cite{WDHJB99,HKLHFC08,DCNG05}.  This QS-based relaying model allows
to achieve an arbitrarily long-distance QKD. The idea is roughly
described as follows. One first incrementally build a more long
distance EPR pair from two less long distance EPR pairs by a number of
complicated quantum operations as entanglement purification,
entanglement swapping, etc.  The goal of this step is to create EPR
pairs shared between the two target nodes (origin and destination)
that are in an arbitrarily long distance far away. Then, these two nodes
could do an entanglement-based BB84 protocol to establish the secret
key. Besides the capacity over arbitrarily long distances, another
advantage of the QS-based relaying model is that this model allows to
effectively detect eavesdropping at relaying nodes. Indeed, this
model could be considered as untrusted-model.

\subsubsection{Motivation.}
Although the QS-based relaying model gives a very beautiful result in
theory, working on entangled photons is not easy in practice. In
compared with single-photon approaches, entanglement-based ones seem
to be surcharged by the quickly unavoidable decoherence of entangled photons
over transmission and in time. This fact encourages us looking
for new relaying methods that restrict the use of entangled photons,
or at least effectively decrease the time conserving entangled
photons to get more easy in practical implementations.

Therefore, we first propose the Quantum Quasi-Trusted Bridge
(QQTB) model that is capable of doubling the limited QKD range without
invoking entanglements. Then, we propose the Quantum Quasi-Trusted
Relay (QQTR) model that could be considered as an extended-QQTB
version. The QQTR model is capable of securely distributing shared
keys over arbitrarily long distances. This model works with entangled
photons, but does not need invoke entanglement swapping and
entanglement purification as proposed in
\cite{WDHJB99,HKLHFC08,DCNG05}. As a result, we could effectively
decrease the time required for conserving the coherence of
entangled-photons in compared to previous works. However, the
most originality of our works is two untrusted models: the Quantum
Untrusted Bridge (QUB) and Quantum Untrusted Relay (QUR) ones. The QUB
and QUR models have the same capacity with the QQTB and QQTR ones in the range
point of view, respectively. However, the QUB and QUR models allow to deal
with untrusted intermediate nodes as models proposed in \cite{WDHJB99,HKLHFC08,DCNG05}. That means that the origin and the
destination could effective detect eavesdropping even though attackers
have the full controls over some intermediate nodes.

\section{\label{sec:bck_grd}Background}
We remind some background concepts and make some propositions that are used
to build our four models in the rest of this paper.
 
\subsection{The controlled-NOT (C-NOT) gate}

\begin{figure}[htbp]
  \begin{center}
    \scalebox{0.9}{
      $$
      \input{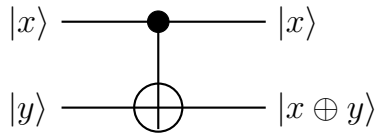} $$}
  \caption{The two-qubit controlled-NOT (C-NOT) gate, also called
    the XOR gate.}
    \label{fig:cnot_gate}
  \end{center}
\end{figure}

Our models need use the quantum controlled-NOT (C-NOT) gate (see Fig.\ref{fig:cnot_gate}). The BB84
protocol does not need use this gate, however, entanglement swapping
and entanglement purification operation require this gate \cite{AVS05,KFHM06}. In fact, the C-NOT gate
is one of the most popular two-qubit quantum gates
\cite{AE00,MNIC00}. Without loss of generality, we will work only
with the C-NOT gate that operates in basis $|+\rangle$ with two
corresponding basis states $|0\rangle$ and $|1\rangle$. By definition,
this gate flips the second (target) qubit if the first (control) qubit
is $|1\rangle$ and does nothing if the control qubit is $|0\rangle$.

\begin{proposition}
\label{propos:same_basis}
If two input qubits are basis states of one sole basis, then:
\begin{enumerate}
\item For the case of the input basis being $|+\rangle$, the XOR of
  two input qubits appears at the second output (exactly as described
  in Fig. \ref{fig:cnot_gate}).
\item For the case of the input basis being $|\times\rangle$, the XOR
  of two input qubits appears at the first output (not as described in
  Fig. \ref{fig:cnot_gate}).
\end{enumerate}
\end{proposition}

\textbf{\textsl{Proof:}} The two basis states of the basis $|+\rangle$
are $|0\rangle$ and $|1\rangle$, corresponding to logical values $0_L$
and $1_L$, respectively. Consequently, the two basis states of the
basis $|\times\rangle$ are $\frac{|0\rangle+|1\rangle}{\sqrt{2}}$ and
$\frac{|0\rangle-|1\rangle}{\sqrt{2}}$, corresponding to logical
values $0_L$ and $1_L$, respectively. As mentioned above, the C-NOT 
gate operates in basis $|+\rangle$.

If the two input qubits are in basis $|+\rangle$, then by the definition
of the C-NOT gate we have:
\begin{equation*}
  \begin{split}
    |0_L\rangle . |0_L\rangle =|0\rangle . |0\rangle & \mapsto_{CNOT} |0\rangle . |0\rangle =|0_L\rangle . |0_L\rangle\\
    |0_L\rangle . |1_L\rangle =|0\rangle . |1\rangle & \mapsto_{CNOT} |0\rangle . |1\rangle =|0_L\rangle . |1_L\rangle\\
    |1_L\rangle . |0_L\rangle =|1\rangle . |0\rangle & \mapsto_{CNOT} |1\rangle . |1\rangle =|1_L\rangle . |1_L\rangle\\
    |1_L\rangle . |1_L\rangle =|1\rangle . |1\rangle & \mapsto_{CNOT} |1\rangle . |0\rangle =|1_L\rangle . |0_L\rangle
  \end{split}
\end{equation*} 

Obviously, the XOR appears at the second output for the case of the
input basis being $|+\rangle$, exactly as described in
Fig.~\ref{fig:cnot_gate}.

We now observe the case in which two input qubits are in basis
$|\times\rangle$.

\begin{equation*}
  \begin{split}
    |0_L\rangle . |0_L\rangle =
    \frac{|0\rangle+|1\rangle}{\sqrt{2}}.\frac{|0\rangle+|1\rangle}{\sqrt{2}}\mapsto & \frac{1}{2}(|0\rangle .(|0\rangle+|1\rangle) +|1\rangle.(|1\rangle+|0\rangle)) = |0_L\rangle . |0_L\rangle\\
    |1_L\rangle . |0_L\rangle =
    \frac{|0\rangle-|1\rangle}{\sqrt{2}}.\frac{|0\rangle+|1\rangle}{\sqrt{2}}
    \mapsto & \frac{1}{2}(|0\rangle . (|0\rangle+|1\rangle)
    -|1\rangle.(|1\rangle+|0\rangle)) = |1_L\rangle . |0_L\rangle\\
    |0_L\rangle . |1_L\rangle = \frac{|0\rangle+|1\rangle}{\sqrt{2}}.\frac{|0\rangle-|1\rangle}{\sqrt{2}}\mapsto &\frac{1}{2}(|0\rangle .(|0\rangle-|1\rangle)+|1\rangle.(|1\rangle-|0\rangle))=|1_L\rangle. |1_L\rangle\\
    |1_L\rangle.|1_L\rangle=\frac{|0\rangle-|1\rangle}{\sqrt{2}}.\frac{|0\rangle-|1\rangle}{\sqrt{2}}\mapsto
    & \frac{1}{2}(|0\rangle .(|0\rangle-|1\rangle) -|1\rangle.(|1\rangle-|0\rangle)) = |0_L\rangle . |1_L\rangle
  \end{split}
\end{equation*}

We realize that the C-NOT gate now changes the roles of two input
qubits. If the second qubit is $1_L$ (in basis $|\times\rangle$) then
it flips the first qubit (in basis $|\times\rangle$). Otherwise, it
does nothing. The XOR (in basis $|\times\rangle$) is at the first
output in this case, not as described in Fig.~\ref{fig:cnot_gate}

\begin{proposition}
\label{propos:dif_basis}
If the two input qubits of the C-NOT gate are basis states in the
different basis (one in $|+\rangle$ and other in $|\times\rangle$),
then
\begin{enumerate}
\item If the first and second qubits are basis states in basis
  $|\times\rangle$ and $|+\rangle$, respectively, then the output is
  an entanglement.
\item If the first and second qubits are basis states in basis
  $|+\rangle$ and $|\times\rangle$, respectively, then the C-NOT gate
  does nothing.
\end{enumerate}
\end{proposition}

\textbf{\textsl{Proof:}} If the first (control) and second (target)
qubits are the basis states in basis $|\times\rangle$ and $|+\rangle$,
respectively, then we have:

\begin{equation*}
  \begin{split}
    \frac{|0\rangle+|1\rangle}{\sqrt{2}}.|0\rangle\mapsto
    \frac{1}{\sqrt{2}}(|0\rangle |0\rangle +|1\rangle |1\rangle) &
    \frac{|0\rangle-|1\rangle}{\sqrt{2}}.|0\rangle\mapsto
    \frac{1}{\sqrt{2}}(|0\rangle |0\rangle -|1\rangle |1\rangle)\\
    \frac{|0\rangle+|1\rangle}{\sqrt{2}}.|1\rangle\mapsto
    \frac{1}{\sqrt{2}}(|0\rangle |1\rangle +|1\rangle |0\rangle) & 
    \frac{|0\rangle-|1\rangle}{\sqrt{2}}.|1\rangle\mapsto
    \frac{1}{\sqrt{2}}(|0\rangle |1\rangle -|1\rangle |0\rangle)
  \end{split}
\end{equation*}

Obviously, the output is an entanglement, more precisely, one of four
Bell states if the first and second qubits are in basis
$|\times\rangle$ and $|+\rangle$, respectively.

If the control and target qubits are the basis states in basis
$|+\rangle$ and $|\times\rangle$, respectively, then we have:
\begin{equation*}
\begin{split}
|0\rangle.\frac{|0\rangle+|1\rangle}{\sqrt{2}}\mapsto &
\frac{1}{\sqrt{2}}(|0\rangle|0\rangle+|0\rangle|1\rangle)=
|0\rangle.\frac{|0\rangle+|1\rangle}{\sqrt{2}}\\
|0\rangle.\frac{|0\rangle-|1\rangle}{\sqrt{2}}\mapsto &
\frac{1}{\sqrt{2}}(|0\rangle|0\rangle-|0\rangle|1\rangle)=
|0\rangle.\frac{|0\rangle-|1\rangle}{\sqrt{2}}\\
|1\rangle.\frac{|0\rangle+|1\rangle}{\sqrt{2}}\mapsto &
\frac{1}{\sqrt{2}}(|1\rangle|1\rangle+|1\rangle|0\rangle)=
|1\rangle.\frac{|0\rangle+|1\rangle}{\sqrt{2}}\\
|1\rangle.\frac{|0\rangle-|1\rangle}{\sqrt{2}}\mapsto &
\frac{1}{\sqrt{2}}(|1\rangle|1\rangle-|1\rangle|0\rangle)=
-|1\rangle.\frac{|0\rangle-|1\rangle}{\sqrt{2}}
=|1\rangle.\frac{|0\rangle-|1\rangle}{\sqrt{2}}
\end{split}
\end{equation*}

Obviously, the C-NOT does nothing in this case.

\subsection{A simple quantum circuit}
\begin{figure}[htbp]
  \begin{center}
    \scalebox{0.9}{
      $$
      \input{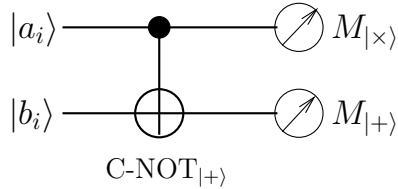} $$}
  \caption{The pair ($|a_i\rangle,|b_i\rangle$) passes through a C-NOT gate
    operating in basis $|+\rangle$ before being measured
    independently: the first output qubit is measured in basis
    $|\times\rangle$ and the second one is measured in basis
    $|+\rangle$.}
    \label{fig:cnot_m_circuit}
  \end{center}
\end{figure}

In fact, we use the C-NOT gate to build a simple quantum circuit as
described in Fig. \ref{fig:cnot_m_circuit}. It has two inputs and two
outputs. The two input qubits first pass through a C-NOT gate
operating in basis $|+\rangle$, and then are measured independently by
two quantum detectors that operate in different basis $|\times\rangle$
and $|+\rangle$ (see Fig. \ref{fig:cnot_m_circuit}). The two outputs
are classical bits $0$ or $1$. We can directly deduce from Proposition
\ref{propos:same_basis} to the following proposition.

\begin{proposition}
  \label{propos:cnot_m_circuit}
  If two input qubits are basis states of one sole basis, then the
  quantum circuit as described in Fig. \ref{fig:cnot_m_circuit} does
  an irreversible XOR operation. This circuit reveals no more
  information than the logical XOR of two logical inputs. Indeed,
  \begin{enumerate} 
  \item If two input qubits $|a\rangle$ and $|b\rangle$ are in the
    same basis $|+\rangle$, then the second output is $(a_L\oplus
    b_L)$ and the first output is either $0$ or $1$ with equal
    probabilities ($50\%$ for each one), where $a_L$ and $b_L$ are
    logical values of states $|a\rangle$ and $|b\rangle$ in basis
    $|+\rangle$, respectively.
  \item If two input qubits $|a\rangle$ and $|b\rangle$ are in the
    same basis $|\times\rangle$ then the first output is $(a_L\oplus
    b_L)$ and the second output is either $0$ or $1$ with equal
    probabilities ($50\%$ for each one), where $a_L$ and $b_L$ are
    logical values of states $|a\rangle$ and $|b\rangle$ in basis
    $|\times\rangle$, respectively.
  \end{enumerate}
\end{proposition}

\subsection{EPR pairs - Bell states}
A Bell state (or an EPR pair) is defined as a maximally entangled
quantum state of two qubits. These qubits could be spatially
separated, however, they always exhibit perfect correlations. Assume
that Alice and Bob share one of four Bell states
$|\Phi^+\rangle=\frac{1}{\sqrt{2}}(|\uparrow_A\uparrow_B\rangle +
|\downarrow_A\downarrow_B\rangle)$. If Alice and Bob measure their
qubits in any common basis at their spatially separated sites, then Alice will get a random logical
output either $0_L$ or $1_L$ with each probability of $50\%$ and the
output of Bob is always parallel with that of Alice (i.e. the same value).

If we focus on the logical values then we could describe
four Bell states that form an orthogonal basis for the quantum state
of two qubits as follows:

\begin{equation*}
  |\Phi^\pm\rangle = |0_L0_L\rangle \pm |1_L1_L\rangle \qquad |\Psi^\pm\rangle = |0_L1_L\rangle \pm |1_L0_L\rangle
\end{equation*}

\section{\label{sec:qqtr_def}Quantum Quasi-Trusted (QQT) Relays}
Let us observe a three-party communication scenario as follows. The origin Alice wants to
establish a secret key with the destination Bob. They want to achieve 
unconditional security. However, the distance between them exceeds the
limited range of QKD. Carol is a intermediate node that could share
QKD links with Alice and Bob. It seems reasonable to assume that Carol is
honest enough to correctly follow a given three-party communication
protocol even though she is eavesdropped by the malicious person Eve. In
such a scenario, we call Carol a \textsl{quasi-trusted} relay.

\begin{definition}[QQT relay]
\label{def:QQT_relay}
A Quantum Quasi-Trusted (QQT) relay is a person or a station that can
execute simple quantum operations as measurement, C-NOT, etc., and
holds the following conditions :
\begin{enumerate}
\item \textsl{Finite-Time Trust:} The relay is honest enough to
  correctly implement a given finite-time communication
  protocol. After having done the protocol, the relay could be
  corrupted.
\item \textsl{Under eavesdropping:} The relay could be always under
  the monitoring of eavesdroppers.
\end{enumerate} 
\end{definition}

Our quasi-trusted relay definition is simple but very important
since we will use it to build the Quantum Quasi-Trusted Bridge
(QQTB) and Quantum Quasi-Trusted Relay (QQTR) models in the next of
this paper. 

\section{\label{sec:qqtb_model}Quantum Quasi-Trusted Bridge (QQTB) model}
\subsection{Description}
\begin{figure}[hbtp]
  \centerline{\includegraphics[width=0.8\linewidth]{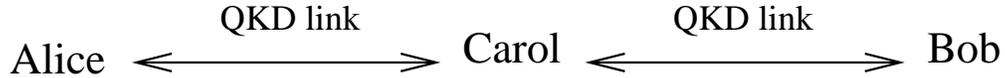}}
  \caption{QKD bridge: Alice and Bob are out of the QKD range, they
    want to use Carol as a bridge to communicate securely the shared
    key.}
  \label{fig:qkd_bridge}
\end{figure}

\begin{definition}[QQTB model]
\label{def:QQTB_model}
The QQT-bridge (QQTB) model is a three-party communication model in
which the QQT relay Carol acts as a bridge that helps two long-distance nodes Alice and Bob
to securely establish a shared key. The Fig. \ref{fig:qkd_bridge}
roughly describes a QQTB model.
\end{definition}

The QQTB model uses an implicit assumption that Eve cannot eavesdrop
the origin Alice and the destination Bob. Such a assumption is trivial
since if the origin or the destination is eavesdropped then there is
no solution for Alice and Bob. Our definition of the QQTB model also implies that Eve is allowed to
execute classical and quantum attacks over the channels Alice-Carol
and Carol-Bob, even over Carol's
site. At the first glance, we realize that the most dangerous
vulnerability is from the Carol's site. Indeed, although the two
channels Alice-Carol and Carol-Bob could be secured by QKD (see Fig
\ref{fig:qkd_bridge}), if information appears in clear at the Carol's site then
Eve could get it without leaving any trace by eavesdropping.

\subsection{Protocols}
The problem is how we could design secure three-party
communication protocols that satisfy the constraints of the QQTB
model, implicitly, that hold the conditions of the QQT relay (see
Definition \ref{def:QQT_relay}). We develop a simple idea that is
based on the one-time pad unbreakable encryption scheme. The idea
could be described as follows. We try to create the situation in which Alice,
Carol and Bob own three pads $A, C, B$, respectively. These pads hold
$C=A \oplus B$ (a bit-wise XOR operation). Note that Carol owns $C$
and knows no more than $C=A \oplus B$. When Alice wants to send to
Bob a secret key $K$, she sends $K \oplus A$ to Carol. Carol
receives $K \oplus A$, computes $K \oplus A \oplus C = K
\oplus B$, and sends the result to Bob. Bob receives $K \oplus
B$, computes $K \oplus B \oplus B$ to obtain $K$.  In such a
situation, even though Carol owns $C = A \oplus B$, she cannot
reveal $K$. Besides, the key $K$ is unconditional secured
over channel since we use the one-time pad scheme. Obviously, Carol holds the \textsl{under eavesdropping}
condition (see Definition
\ref{def:QQT_relay}) and could be developed to become a QQT bridge. 

We will begin with a classical protocol that illustrates our approach.
This protocol is not secure. Then we turn into quantum world to see
how quantum mechanics could help.

\subsubsection{The Insecure Quasi-Trusted Bridge (IQTB) protocol:} The
protocol consists of the following steps.
\begin{enumerate}
\item Alice securely sends to Carol a random $m$-bit string $A$ by a QKD link.
\item Bob securely sends to Carol a random $m$-bit string $B$ by a QKD
  link.
\item Carol receives $A$ and $B$, computes $C=A\oplus B$ (XOR
  operation).
\item \textsl{Carol deletes $A$ and $B$ in her memory device.}
\item Transmitting the secret key:
  \begin{itemize} 
  \item Alice randomly creates the $m$-bit key $K$, sends $K\oplus
    A$ to Carol. 
  \item Carol receives $K\oplus A$, computes $K\oplus A\oplus
    C = K\oplus B$, then sends the result to Bob.
  \item Bob receives  $K\oplus B$, computes $K\oplus B \oplus
    B$ to obtain $K$.
  \end{itemize}
\end{enumerate}

What is insecure in this protocol? The step 4 seems helpful in face
with the \textsl{finite-time trusted} condition of the quasi-trusted
bridge: after having done the protocol, even though Carol is corrupted
the key $K$ is not compromised. But this is not so! Nobody can sure
that in the one hand Carol still does correctly the protocol but in
the other hand she makes copies of $A$ and $B$, maybe only for her curiousness
purpose. And then, when the protocol has been yet finished, she could
be corrupted and gives these copies to Eve. Consequently, the key $K$
is compromised. More seriously, the protocol cannot hold the
\textsl{under-eavesdropping} condition (see Definition
\ref{def:QQT_relay}). Indeed, if Eve could monitor Carol's memory
devices, then she can make herself copies of $A$ and $B$. If $A$ or $B$
is compromised then the key $K$ is compromised, consequently.

Now, we propose the Quantum Quasi-Trusted Bridge (QQTB) protocols that really help Alice and Bob
to securely establish the shared key $K$ through the bridge
Carol. These protocols could defeat drawbacks of the previous protocol. 

\subsubsection{The Quantum Quasi-Trusted Bridge (QQTB) Protocol:} The
protocol consists of 4 main steps.

\textbf{Step 1:} Preparing, exchanging, and measuring qubits.
\begin{enumerate}
\item Alice creates $2n$ random bits and chooses the random $2n$-bit
  string $b_A$. For each bit $i$, she creates a state in basis
  $|+\rangle$ or $|\times\rangle$ if $b_A[i]=0$ or $b_A[i]=1$,
  respectively. Alice sends the resulting qubits
  $|a_1,a_2,..,a_{2n}\rangle$ to Carol.
\item Similarly, Bob creates a random $2n$-bit value, the random
  $2n$-bit string $b_B$ and the corresponding qubits
  $|b_1,b_2,..,b_{2n}\rangle$. Then, he sends
  $|b_1,b_2,..,b_{2n}\rangle$ to Carol.
\item Carol receives two $2n$-qubit strings from Alice and Bob in a
  synchronous manner. It means that she receives one by one for all
  the $2n$ pairs ($|a_i\rangle, |b_i\rangle)$. On the arrival of a
  pair, Carol randomly turns into either Check-Mode (CM) or Message-Mode (MM).
  \begin{itemize}
  \item In the CM, Carol measures independently both $|a_i\rangle$ and
    $|b_i\rangle$ in a randomly chosen basis $|+\rangle$ or
    $|\times\rangle$. She gathers both two resulting bits and keeps track
    of their corresponding basis. Note that in this mode Carol does not use the
    quantum circuit described as Fig. \ref{fig:cnot_m_circuit}.
  \item In the MM, Carol first leads $|a_i\rangle, |b_i\rangle$ to two
    inputs of a C-NOT gate, and then measures the two outputs in two
    different basis: the first one in $|\times\rangle$ and the second
    one in $|+\rangle$ as described in Fig. \ref{fig:cnot_m_circuit}.
    She randomly chooses one out of two outputs to keep the measured value and
    the corresponding basis. She discards the another one.
  \end{itemize}
  At the end of the receiving process, the CM and MM's choices roughly
  result in two $n$-position strings: the check-position string $CP=
  cp_1,.., cp_{n}$ and the message-position string $MP=
  mp_1,..,mp_{n}$.
\end{enumerate}

\textbf{Step 2:} Checking for the presence of Eve.
\begin{enumerate}
\item For the channel between Alice and Carol: Alice and Carol communicate their
  basis used in the check-positions $CB$ and the corresponding values.
  They discard positions where their basis are different. They compare
  values at remaining positions.  If some of these values agree, they
  conclude that the channel was compromised. In this case, they
  inform to Bob to abort the whole protocol.
\item For the channel between Bob and Carol: Bob and Carol communicate their
  basis used in the check-positions $CB$ and the corresponding values.
  They discard positions where their basis are different. They compare
  values at remaining positions.  If some of these values agree, they
  conclude that the channel was compromised. In this case, they
  inform to Alice to abort the whole protocol.
\end{enumerate}

\textbf{Step 3:} Creating the pads for Alice, Carol and Bob.
\begin{enumerate}
\item Alice, Carol and Bob announce their basis used in positions $MB=
  mp_1,..,mp_{n}$.
\item If their basis are different at $mp_i$, then they discard this
  position and the corresponding values.
\item The values of the remaining positions result in three pads $A=
  A_1,..,A_m ; C = C_1,..,C_m ; B=B_1,..,B_m$ for Alice, Carol and
  Bob, respectively. These pads hold $C_i = A_i \oplus B_i,
  i\in[1,..,m], m\sim \frac{n}{4}$.
\end{enumerate}

\textbf{Step 4:} Transmitting the key $K$.
\begin{enumerate}
\item Alice creates the random $m$-bit key $K$. She sends $K \oplus
  A$ to Carol.
\item Carol receives from Alice $K \oplus A$, does a XOR operation
  over it and her pad, then sends the result to Bob. Since $C= A
  \oplus B$, the result of the XOR operation is $K \oplus A
  \oplus C = K \oplus B$.
\item Bob receives $K \oplus B$, computes $K \oplus B \oplus
  B$ to obtain $K$.
\end{enumerate}

We show now why this protocol is secure. At the step 1, when a pair
$(|a_i\rangle,|b_i\rangle)$ synchronously arrives to Carol, she
randomly turns into either the Check-Mode (CM) or the Message-Mode (MM). Since
Eve does not know in advance the choices of Carol, she cannot treat
differently the pairs $(|a_i\rangle,|b_i\rangle)$. Therefore, the
error-rate on the check bits must behave like that on the message
bits. In the other hand, the error-check procedures in the channels
(Alice, Carol) and (Carol, Bob) are done exactly as that of the
BB84 protocol. By that, the QQTB protocol's security is exactly the
security of the BB84 protocol. This implies that the QQTB protocol is
unconditionally secure. Readers being interested in security
proof of BB84 are invited to read \cite{PSJP00,HKL01,Chau02,MAYER01}.

One can claim that Carol could unintentionally select some choices of CM
or MM before arrivals of $(|a_i\rangle,|b_i\rangle)$. If Eve knows these choices by
eavesdropping, then she avoids the pairs in CM and attacks on the pairs in
MM. This makes security compromised. 
We propose another protocol that can tolerate such a mistake of Carol.

\subsubsection{The modified-QQTB Protocol:} The protocol consists of 5
main steps.

\textbf{Step 1:} Preparing, exchanging, and measuring qubits.
\begin{enumerate}
\item Alice creates $2n$ random bits and chooses the random $2n$-bit
  string $b_A$. For each bit $i$, she creates a state in a basis
  $|+\rangle$ or $|\times\rangle$ for $b_A[i]=0$ or $b_A[i]=1$,
  respectively. Alice sends the resulting qubits
  $|a_1,a_2,..,a_{2n}\rangle$ to Carol.
\item Similarly, Bob creates a random $2n$-bit value, the random
  $2n$-bit string $b_B$ and the corresponding qubits
  $|b_1,b_2,..,b_{2n}\rangle$. Then, he sends
  $|b_1,b_2,..,b_{2n}\rangle$ to Carol.
\item Carol receives two $2n$-qubit strings from Alice and Bob in a
  synchronous manner. It means that she receives one by one for $2n$
  pairs ($|a_i\rangle, |b_i\rangle)$. For each pair, Carol first
  leads $|a_i\rangle, |b_i\rangle$ to the C-NOT gate and then measures
  two output qubits in two different basis: the first one in
  $|\times\rangle$ and the second one in $|+\rangle$ as described in
  Fig. \ref{fig:cnot_m_circuit}. Then, she randomly chooses one out of two
  outputs to keep the measured value and the corresponding basis.
  She discards the other one.
\end{enumerate}

\textbf{Step 2:} Sifting.
\begin{enumerate}
\item Alice, Carol and Bob announce their basis.
\item If their basis are different at the position $i$, then they discard this position.
\item The values of the remaining positions
  result in three $2m$-bit strings $a= a_1, ..,a_{2m} ; c = c_1,.., c_{2m} ;
  b=b_1,..,b_{2m}$ for Alice, Carol and Bob, respectively. Theoretically, these three
  strings hold $c_i = a_i \oplus b_i, i\in[1,2m], 2m\sim \frac{2n}{4}$.
\end{enumerate}

\textbf{Step 3:} Checking for the presence of Eve.
\begin{enumerate}
\item Alice, Carol, Bob randomly agree $m$ out of $2m$ positions to
  check the presence of Eve. This results in two $m$-position strings:
  the check-position string $CP= cp_1,..,cp_{m}$ and the
  message-position string $MP= mp_1,..,mp_{m}$.
\item Alice, Carol, Bob announce their values $a_{cp_i},
  b_{cp_i}, c_{cp_i}$, respectively, in check-positions $cp_i$. They
  check if $c_{cp_i}=a_{cp_i} \oplus b_{cp_i}$ or not. If some of negative checks,
  they abort the protocol. 
\end{enumerate}

\textbf{Step 4:} Creating the pads for Alice, Carol and Bob.
\begin{enumerate}
\item The values in $m$ message-positions result in three $m$-bit
  pads $A= A_1,..,A_m ; C = C_1,.., C_m ; B=B_1,..,B_m$ for Alice,
  Carol and Bob, respectively. These pads hold $C_i = A_i \oplus
  B_i, i\in[1,..,m], m\sim \frac{n}{4}$.
\end{enumerate}

\textbf{Step 5:} Transmitting the key $K$.
\begin{enumerate}
\item Alice creates the random $m$-bit key $K$. She sends $K
  \oplus A$ to Carol.
\item Carol receives $K \oplus A$ from Alice, does a XOR operation
  over it and her pad, then sends the result to Bob. Since $C= A
  \oplus B$, the result of the XOR operation is $K \oplus A
  \oplus C = K \oplus B$.
\item Bob receives $K \oplus B$, computes $K \oplus B \oplus
  B$ to obtain $K$.
\end{enumerate}

This protocol makes sure that measurements are done before check-position and message-position choices. The classical information that
could be eavesdropped by Eve on the
Carol site now does not reveal any
information of $K$. We must show that the protocol is
secure in faced against Eve's attacks over channel. From our three-party
communication model, we build a virtual two-party
communication between Anna and Borris in which:
\begin{enumerate}
\item Anna plays the roles of both Alice and Bob.
\item Borris plays the role of Carol.
\item The virtual channel between Anna and Borris consists of both two
  real channels (Alice, Carol) and (Carol, Bob).
\end{enumerate}

Let Anna and Borris do our modified QQTB protocol. We realize that
Anna and Borris do a variant of the BB84 protocol that takes the same
principles. Anna codes a classical bit by non-orthogonal quantum
states $|q_1\rangle|q_2\rangle$, where $|q_1\rangle,|q_2\rangle$ are simultaneously
in basis $|+\rangle$ or $|\times\rangle$. Borris receives a
classical bit by measuring $|q_1\oplus
q_2\rangle$ in a random basis $|+\rangle$ or $|\times\rangle$. If his basis choice
is right then the receiving value is exactly $q_1\oplus q_2$.
 Otherwise, the receiving bit has a probability of $50\%$ to be
right. Eve cannot attack such a conjugate code without
introducing more disturbances over channel. By estimating the disturbance, we
could detect the presence of Eve over the virtual channel.
 This implies that we can make sure either the channels
(Alice, Carol) and (Carol, Bob) are attacked or not as in the BB84 protocol. In other words, our modified QQTB
protocol is unconditionally secure. Readers being interested in security
proof of BB84 are invited to read \cite{PSJP00,HKL01,Chau02,MAYER01}.

\subsubsection{The enhanced-QQTB Protocol:} In the modified-QQTB, we
realize that if Alice and Bob use a
common basis at the position $i$, then the quantum
circuit at the Carol site gives no more information than the logical XOR
of two logical values of Alice and Bob. Therefore, no need to force
Carol randomly choosing to keep one output (one measuring basis) before Alice
and Bob announcing publicly their basis. The enhanced-QQTB protocol is
very similar to the modified-QQTB ond. But it could improve the
secret-bit rate up to two times.

The enhanced QQTB consists of 5 steps.

\textbf{Step 1:} Preparing, exchanging, and measuring qubits.
\begin{itemize}
\item Alice, Carol, Bob do as in the modified QQTB protocol. However,
  instead of keeping only one output, Carol keeps informations of both
  two outputs.
\end{itemize}

\textbf{Step 2:} Sifting.
\begin{itemize}
\item Alice and Bob announce their basis: if the basis are different at the position $i$, then Alice,
  Bob, and Carol discard the position $i$.
\item For each remaining position $i$, Carol keeps only informations
  (value and basis) of either
  the first output or the second one if the common basis used by Alice and Bob is $|\times\rangle$ or $|+\rangle$, respectively. She discards
  informations of the other one.
\item Now, the values of the remaining positions
  result in three $2m$-bit strings $a= a_1, ..,a_{2m} ; c = c_1,.., c_{2m} ;
  b=b_1,..,b_{2m}$ for Alice, Carol and Bob, respectively. Theoretically, these three
  strings hold $c_i = a_i \oplus b_i, i\in[1,2m], 2m\sim \frac{2n}{2}$.
\end{itemize}

\textbf{Step 3:} Checking for the presence of Eve.
\begin{itemize}  
\item Alice, Carol, and Bob do exactly as in the modified QQTB protocol.
\end{itemize}

\textbf{Step 4:} Creating the pads for Alice, Carol and Bob.
\begin{itemize}
\item Alice, Carol, and Bob do exactly as in the modified QQTB protocol.
\end{itemize}

\textbf{Step 5:} Transmitting the key $K$.
\begin{itemize}
\item Alice, Carol, and Bob do exactly as in the modified QQTB protocol.
\end{itemize}

Note that the security of enhanced QQTB protocol is exactly that of
the modified QQTB protocol since the quantum circuit at the Carol site
reveals no more than the XOR result and the qubit measurements are
always done before Alice and Bob revealing theirs basis. Eve always
deals with unknown states as in the modified-QQTB protocol. Therefore, the enhanced-QQTB
protocol also gives unconditional security. However, the number of
secret bits obtained from the enhanced QQTB protocol is two time
bigger than that of the modified QQTB protocol (see $m$ in the step 2).

\section{\label{sec:qqtr_model}Quantum Quasi-Trusted Relay (QQTR) model}

\subsection{Is is possible to extend QQTB model over arbitrarily long
  distance?}

The information theory states that Alice and Bob cannot publicly agree a common
secret unless they pre-possess a secret key that has the length at
least equal to that of the secret \cite{CS49}. The quantum mechanic
opens a new door that allows Alice and Bob to achieve their
goal. Indeed, the quantum no-cloning theorem states that it is
impossible to make a perfect copy of un unknown quantum state. This
implies that eavesdropping on quantum channels will introduce some
detectable disturbance. By estimating the error rates, Alice and Bob
can effectively detect the presence of the eavesdropper Eve.  

In this paper, we study quantum models that work
with perfect quantum devices, quantum free-error channels, and without
quantum memory devices. Note that although the quantum channels are
assumed free-error, we should take into account the degradations of single-photon
energy and entangled-photon coherence over transmission. We assume
that our quantum devices are perfect if and only if the
single-photon energy and the entangled-photon coherence are above some
given thresholds. In QQTB model, we implicitly address single
photon schemes to avoid difficulties arising from entanglement
decoherence. The question is whether we could extend this model based
on single photon up to an arbitrarily long distance? We observe the
scenario in which there is Dave in the right of Bob. Now Bob plays the
role of quasi-trusted relay as Carol. The goal is that Alice could
convey a secret to Dave, not to Bob. Assume that the distances between Alice,
Carol, Bob and Dave are the critical distances of single-photon
transmission over that transmitted qubits are correctly detected. In other words, Alice cannot send directly a single
photon to Bob or Dave, and Dave cannot send directly a single photon to
Carol or Alice. Therefore, Alice and Dave cannot make together a quantum contact at one
unique intermediate location as the spirit of the QQTB
model. In the other hand, any classical contact is no help. Therefore, we could conclude that the single-photon based QQTB
model cannot extend more than two time of the limited QKD range. This
makes the senses of the word ``bridge'' in the QQTB model: two bridges
cannot be build successively.

\subsection{Quantum Quasi-Trusted Relay (QQTR) model}
\subsubsection{QQTR model's description.}
We take into account EPR pairs to build our QQTR model. As mentioned
in Section \ref{sec:rw_motiv}, we try to limit the time 
keeping EPR pairs to avoid difficulties arising from entanglement
decoherence. Such a motivation makes our
works distinguished from the works presented in \cite{WDHJB99,HKLHFC08,DCNG05}.

\begin{figure}[htbp]
  \centerline{\includegraphics[width=0.8\linewidth]{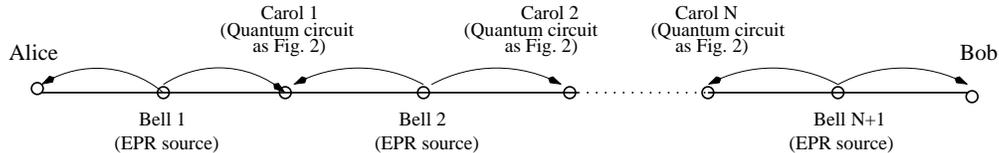}}
  \caption{Bell 1,.., Bell N are EPR-pair sources. Carol
    1, .., Carol N act as Carol in the enhanced-QQTB protocol.}
  \label{fig:qqtr_model}
\end{figure}

The QQTR model is roughly described as Fig.\ref{fig:qqtr_model}.

\subsubsection{QQTR protocol.}
Between Alice and Bob we arrange $N$ Carols ($C_1,..,C_N$ for short)
and $N+1$ Bells ($B_1,..,B_{N+1}$ for short) as described in
Fig.\ref{fig:qqtr_model}. This creates $2N+2$ segments. Without loss of
generality, we assume that the length of segments are the same and
the segment length allows our quantum devices working correctly and
effectively on entanglement coherence and single-photon detection.
 
Our QQTR protocol consists of 5 steps:

\textbf{Step 1:} Preparing, exchanging, and measuring qubits.
\begin{enumerate}
\item For $B_1, .., B_{N+1}$, each prepares $n$ Bell states
  $(|\Phi^+\rangle)^n$.
\item $B_1$ sends the first half of each Bell state to Alice (the previous), the
  second half to $C_1$ (the next). $B_{N+1}$ sends the first half of each Bell
  state to $C_N$ (the previous), the second half to Bob (the next). For $i\in[2,N]$, $B_i$
  sends the first half of each Bell state to $C_{i-1}$ (the previous), the
  second half to $C_i$ (the next).
\item Each $C_i, i\in[1,N],$ receives $2n$ qubits from $B_i$ and
  $B_{i+1}$ in a synchronous manner. This means that she receives $n$
  times, and for each time she leads the qubit from $B_i$ and the
  qubit from $B_{i+1}$ to the first and second inputs of the quantum
  circuit as described in Fig. \ref{fig:cnot_m_circuit}. Then, she
  keeps informations (the measured
  value and the corresponding basis) of both two outputs. Briefly, $C_i$ acts exactly as
  Carol in the enhanced-QQTB protocol.
\item Alice and Bob receive $n$ qubits for each one. They randomly
  choose basis to measure their qubits, independently.
\end{enumerate}

\textbf{Step 2:} Sifting.
\begin{enumerate}
\item Alice and Bob announce their $n$ basis used.
\item If the basis are different at the position $i$, then Alice,
  Bob, $C_1,..,C_N$ discard this position.
\item For each remaining position $i$, $C_1,..,C_N$ keep only
  informations (value and basis) of either
  the first output or the second output if the common basis of Alice and Bob
  is $|\times\rangle$ or $|+\rangle$, respectively. They discard
  informations of the other one.
\item The values of the remaining positions result in $N+2$ $2m$-bit
  strings $a= a_1, ..,a_{2m} ; c(i) = c(i)_1,.., c(i)_{2m}, i=1..N ;
  b=b_1,..,b_{2m}$ for Alice, $C_1,..,C_N$, and Bob, respectively. These
  $N+2$ strings should hold $\bigoplus_{j=1}^N c(i)_j = a_i \oplus
  b_i, i\in[1,2m], 2m\sim \frac{n}{2}$.
\end{enumerate}

\textbf{Step 3:} Checking for the presence of Eve.
\begin{enumerate}
\item Alice, Bob, and $C_1,..,C_N$ randomly agree $m$ out of $2m$
  positions to check the presence of Eve. This results in two
  $m$-position strings: the check-position string $CP= cp_1,..,cp_{m}$
  and the message-position string $MP= mp_1,..,mp_{m}$.
\item Alice, Bob, $C_1,..,C_N$ announce values in check-position $a= a_{cp_1}, ..,a_{cp_m};
  b=b_{cp_1},..,b_{cp_m} ; c(i) = c(i)_{cp_1},.., c(i)_{cp_m}, i=1..N
  $, respectively. They check if $\bigoplus_{i=1}^N c(i)_{cp_j}=a_{cp_j}
  \oplus b_{cp_j}$ or not. If some of negative checks, they abort
  the protocol.
\end{enumerate}

\textbf{Step 4:} Creating the pads for Alice, $C_1,..,C_N$, and Bob.
\begin{enumerate}
\item The values in $m$ message-positions result in $N+2$ $m$-bit pads
  $P^A= P^A_1,..,P^A_m$; $P^{C(i)} = P^{C(i)}_1,.., P^{C(i)}_m, i
  \in[1,N]$; and $P^B=P^B_1,..,P^B_m$ for Alice, $C_1, .., C_N$, and Bob,
  respectively. These pads hold $\bigoplus_{i=1}^N P^{C(i)} = P^A
  \oplus P^B$.
\end{enumerate}

\textbf{Step 5:} Transmitting the key $K$.
\begin{enumerate}
\item Alice creates the random $m$-bit key $K$, $ m\sim
  \frac{n}{4}$. She sends $K \oplus P^A$ to $C_1$.
\item For $i$ from $1$ to $N-1$, $C_i$ receives $K \oplus P^A
  \oplus \bigoplus_{j=1}^{i-1} P^{C(j)}$, does a XOR operation over
  it and her pad, then sends the result to $C_{i+1}$.
\item $C_N$ receives $K \oplus P^A \oplus \bigoplus_{j=1}^{N-1}
  P^{C(j)}$, does a XOR operation over it and her pad, then sends the
  result to Bob
\item Bob receives $K \oplus P^A \oplus \bigoplus_{j=1}^{N}
  P^{C(j)}$, computes $K \oplus P^A \oplus P^B \oplus
  \bigoplus_{j=1}^{N} P^{C(j)} = K$.
\end{enumerate}

\subsection{Correctness and security}
\subsubsection{Correctness.}
One could claim that is it true that $\bigoplus_{j=1}^N c(i)_j = a_i
\oplus b_i, i\in[1,2m], 2m\sim \frac{n}{2}$ in the step 2
(sifting) of the QQTR protocol?

We first look at 5 sites: Alice, $B_1,C_1,B_2, C_2$. We focus on the
effect of the quantum circuit (see Fig. \ref{fig:cnot_m_circuit}) on
the site $C_1$. This circuit acts on
$|\Phi^+\rangle_{1,2},|\Phi^+\rangle_{3,4}$ coming from $B_1,B_2$.
The subscripts stand for the particle (qubit) numbering. Without
loss of generality, we assume that after having discarded positions of
different basis, the common basis is $|+\rangle$. This implies that
$C_1$, as all $C_2,..,C_N$, keep the result of the second output, and 
discard the first output of the quantum circuit. When the qubits $2,3$
go through the C-NOT gate, we have:

\begin{equation*}
  \begin{split}
    & |\Phi^+\rangle_{1,2}|\Phi^+\rangle_{3,4} = \frac{1}{2}(|0000\rangle_{1234} +|0011\rangle_{1234} + |1100\rangle_{1234} + |1111\rangle_{1234})\\
    & \quad \mapsto_{CNOT_{2,3}} \frac{1}{2}(|0\rangle_1|00\rangle_{23}|0\rangle_4 +|0\rangle_1|01\rangle_{23}|1\rangle_4 + |1\rangle_1|11\rangle_{23}|0\rangle_4 + |1\rangle_1|10\rangle_{23}|1\rangle_4) 
  \end{split}
\end{equation*}

The qubits $2,3$ are measured in basis $|\times\rangle,
|+\rangle$, respectively. This makes the states $|\Phi^+\rangle_{1,2},|\Phi^+\rangle_{3,4}$ collapsed
into a mixed state either $\frac{1}{2}(|00\rangle\langle
00|+|11\rangle\langle 11|)$ or $\frac{1}{2}(|01\rangle\langle
01|+|10\rangle\langle 10|)$ depending on the logical value of the
second output being 0 or 1, respectively. Note that if one gets the
logical values of both two outputs then he can know exactly the qubits
$1,4$ being in what state (one out of four pure states
$|00\rangle,|01\rangle, |10\rangle,|11\rangle$. However, the first
output is measured in basis $|\times\rangle$ and results in a random bit. This means that the logical
value of the qubit $2$ is deleted by a quantum manner. $C_1$ gets only the logical XOR
result of the qubits $2,3$ that is capable of tracking the parity of
the qubits $1,4$. Indeed, if the XOR value is $0$ or $1$, the global
state of two qubits
$1,4$ is either $\frac{1}{2}(|00\rangle\langle
00|+|11\rangle\langle 11|)$ or $\frac{1}{2}(|01\rangle\langle
01|+|10\rangle\langle 10|)$ that has the logical parity either $0$ or $1$,
respectively.

Therefore, when the quantum circuit of $C_1$ has finished, we have a
situation that could be described as follows. We denote the qubits $1,4$
by $|a\rangle$ and $|x_1\rangle$, respectively. Alice owns $|a\rangle$. The qubit
$|x_1\rangle$ is transmitted to $C_2$ to enter, as the first input, into the
quantum circuit at $C_2$. $C_1$ owns a classical bit $c(1)$ that hold
$c(1)=a \oplus x_1$ provided that $|a\rangle$ and $|x_1\rangle$ are 
measured afterward in $|+\rangle$. We now observe the quantum circuit at $C_2$.
\begin{equation*}
  \begin{split}
    & |x_1\rangle|\Phi^+\rangle_{5,6} = \frac{1}{\sqrt{2}}(|x_1\rangle|00\rangle_{56} +|x_1\rangle|11\rangle_{56})\\
    & \quad \mapsto_{CNOT_{x_1,5}} \frac{1}{\sqrt{2}}(|x_1\rangle|x_1\rangle_5|0\rangle_6 +|x_1\rangle|x_1+1\rangle_5|1\rangle_6
  \end{split}
\end{equation*}

After measurements are done at the two outputs, we have the following situation. We
denote the remaining half of the EPR pair (the qubit
6) by $|x_2\rangle$. $|x_2\rangle$ is transmitted to $C_3$. $C_1$ owns the classical
value $c(2)$ that holds $c(2)=x_1 \oplus x_2$ since the qubits $5,6$
are parallel. 

The quantum circuits at the sites $C_3$ to $C_N$ do similarly
as that at $C_2$. This results in: $c(i) = x_{i-1} \oplus x_i, i \in
[2,N]$. Note that Bob owns $|x_N\rangle = |b\rangle$. Finally, when Alice and
Bob measure their qubits, we have:

\begin{equation*}
c(1)= a \oplus x_1 ; \quad c(i)= x_{i-1} \oplus x_i \text{ for } i\in
[2,N-1]; \quad c(N)= x_{N-1} \oplus b
\end{equation*}

Obviously, we have $\bigoplus_{i=1}^N c(i) = a \oplus b$.

\subsubsection{Security.}
We distinguish possible attack types of Eve.
\begin{enumerate}
\item Type 1: Quantum attack on sites Bell 1,.., Bell N+1 ($B_1,..,B_{N+1}$).
\item Type 2: Quantum attack on sites Carol 1, .., Carol N ($C_1,.., C_N$).
\item Type 3: Quantum attack on channel. Eve could do quantum attacks on $2n+2$
  segments between Alice and Bob.
\item Type 4: Classical attack, eavesdropping on sites $C_1,..,C_N$.
\end{enumerate} 

The attack Type 1 implies imperfect EPR sources: the qubit pairs could
be entangled with Eve's probes. In \cite{HKLHFC08}, fortunately, Lo
and Chau have proven that we can effectively check perfect EPR sources
by executing random-hashing verification schemes. As a result, we
could conclude that our QQTR protocol is secure faced to this attack type.

As Carol in the enhanced-QQTB protocol, $C_1, .., C_N$ reveal no
information than the XOR results. Their choices of the first or the
second output depend on the randomness of the basis choices of Alice
and Bob. This implies that all the single states
(qubits) in the
channels (attack type 3) and the $C_1,..,C_N$ (attack type 2) are
unknown states for Eve. By the no-cloning theorem, Eve will make additional
disturbances if she attacks on these states. In the step 3 of the
QQTR protocol, we check the presence of Eve as the checking scheme of the
enhanced-QQTB protocol. Therefore, we could conclude that our QQTR protocol is secure
face to the attack types 2 and 3.

Our protocol also is secure with the attack type 4 since the classical
values $a, b$ were not revealed outside of Alice and Bob. The
knowledge of $c(1),.., c(N)$ cannot deduce exactly the values of $a,b$. Here, we can say that the main idea of the QQTR protocol is
exactly that of the single-photon QQTB protocols. This is the
spirit of our ``quasi-trusted'' model.

\section{\label{sec:qub_model}Quantum Untrusted Bridge (QUB) Model}
\subsubsection{Model description.} 
\begin{figure}[hbtp]
  \centerline{\includegraphics[width=0.8\linewidth]{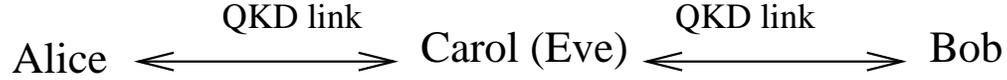}}
  \caption{Alice and Bob are out of the QKD range. They
    must securely transmit shared keys through Eve. This implies that
    they must effectively detect to discard the cases in which Eve
    tries to read the transmitting keys.}
  \label{fig:qub_model}
\end{figure}

The QUB model is very similar to the
QQTB one (see Section \ref{sec:qqtb_model}. However, in this model we
release the ``finite-time trusted'' condition of the intermediate node
Carol. Instead, we require that Alice and Bob must effectively detect
the case in which Carol tries to cheat. This implies that Eve could
have full control on the Carol site or in the other word she plays the
role of Carol (see Fig. \ref{fig:qub_model}). We must design a protocol
that allows Alice and Bob to effectively detect to discard the cases
in which Eve does not correctly follow the protocol and tries to read the transmitting keys.
\subsubsection{The QUB protocol.} The protocol consists of 5 main steps.

\textbf{Step 1:} Preparing, exchanging, and measuring qubits.
\begin{enumerate}
\item Alice creates $2n$ random bits and chooses the random $2n$-bit
  string $b_A$. For each bit $i$, she creates a state in a basis
  $|+\rangle$ or $|\times\rangle$ for $b_A[i]=0$ or $b_A[i]=1$,
  respectively. Alice sends the resulting qubits
  $|a_1,a_2,..,a_{2n}\rangle$ to Carol.
\item Similarly, Bob creates a random $2n$-bit value, the random
  $2n$-bit string $b_B$ and the corresponding qubits
  $|b_1,b_2,..,b_{2n}\rangle$. Then, he sends
  $|b_1,b_2,..,b_{2n}\rangle$ to Carol.
\item Carol receives two $2n$-qubit strings from Alice and Bob in a
  synchronous manner. It means that she receives one by one for $2n$
  pairs ($|a_i\rangle, |b_i\rangle)$. For each pair, Carol first
  leads $|a_i\rangle, |b_i\rangle$ to the C-NOT gate and then measures
  two output qubits in two different basis: the first one in
  $|\times\rangle$ and the second one in $|+\rangle$ as described in
  Fig. \ref{fig:cnot_m_circuit}. 
\item \textsl{Carol sends the values obtained at both two outputs to
    Alice and Bob. The role of Carol stops here.}
\end{enumerate}

\textbf{Step 2:} Sifting.
\begin{enumerate}
\item Alice and Bob communicate their basis.
\item If their basis are different at the position $i$, then they discard this position.
\item At a remaining position $i$, they keep only the second
  output or the first output of Carol for the common basis (at
  position $i$) being $|+\rangle$ or $|\times\rangle$,
  respectively. This reduces a half of the value string came from
  Carol at both the sites of Alice and Bob.
\item Therefore, the values of the remaining positions
  result in three $2m$-bit strings $a= a_1, ..,a_{2m} ; c = c_1,.., c_{2m} ;
  b=b_1,..,b_{2m}$ where Alice keeps two string $a, c$ and Bob keeps
  two strings $b,c$. Theoretically, these three
  strings hold $c_i = a_i \oplus b_i, i\in[1,2m], 2m\sim \frac{2n}{2}$.
\end{enumerate}

\textbf{Step 3:} Checking for the presence of Eve.
\begin{enumerate}
\item Alice, Bob randomly agree $m$ out of $2m$ positions to
  check the presence of Eve. This results in two $m$-position strings:
  the check-position string $CP= cp_1,..,cp_{m}$ and the
  message-position string $MP= mp_1,..,mp_{m}$.
\item Alice, Bob announce their values $a_{cp_i},
  b_{cp_i}$, respectively, in check-positions $cp_i$. They
  check if $c_{cp_i}=a_{cp_i} \oplus b_{cp_i}$ or not. If some of negative checks, they abort the protocol. 
\end{enumerate}

\textbf{Step 4:} Creating the pads for Alice, Bob.
\begin{enumerate}
\item The values in $m$ message-positions result in three $m$-bit
  pads $A= A_1,..,A_m ; C = C_1,.., C_m ; B=B_1,..,B_m$ where Alice
  holds two strings $A, C$ and Bob holds two strings $B,C$. Note that $C_i = A_i \oplus
  B_i, i\in[1,..,m], m\sim \frac{n}{2}$.
\end{enumerate}

\textbf{Step 5:} Transmitting the key $K$.
\begin{enumerate}
\item Alice creates the random $m$-bit key $K$. She sends $K
  \oplus A \oplus C = K \oplus B$ to B.
\item Bob receives $K \oplus B$, computes $K \oplus B \oplus
  B$ to obtain $K$.
\end{enumerate}

\subsubsection{Security.} Note that the quantum circuit of Carol
gives no more information than one XOR result either in basis
$|+\rangle$ or $|\times\rangle$, appearing at the
second output or the first one, depending on the common basis of Alice
and Bob being $|+\rangle$ or $|\times\rangle$, respectively. In the modified-QQTB and
enhanced-QQTB
protocols, since Carol participates in the check process, she could
cheat Alice and Bob. In the QUB protocol, Carol must announce her values
before she knows the choices of basis of Alice and Bob. This implies
that the quantum states of Alice and Bob are really unknown to Carol. If she does not correctly follow the protocol,
then her measured values must introduce some more errors. Note that Carol must always introduce one correct XOR
result of two unknown states came from Alice and Bob, provided the
Alice and Bob's choices of basis is the same. This allows the step 3
of the protocol effectively detect malicious operations of Carol.

\section{\label{sec:qur_model}Quantum Untrusted Relay (QUR) Model}
\subsubsection{Model description.} The QUR model is very similar to the
QQTR one (see Section \ref{sec:qqtr_model}). However, this model
releases the ``finite-time trusted'' condition of the intermediate nodes
Carol. Instead, we require that Alice and Bob must effectively detect to discard the cases
in which Carol does not correctly follow the protocol and tries to
read the transmitting keys. In the other word, the QUR model
works with untrusted intermediate nodes.

\subsubsection{QQTR protocol.}
Between Alice and Bob we arrange $N$ Carols ($C_1,..,C_N$ for short)
and $N+1$ Bells ($B_1,..,B_{N+1}$ for short) as described in
Fig.\ref{fig:qqtr_model}. This creates $2N+2$ segments. Without loss of
generality, we assume that the length of segments are the same and
the segment length allows our quantum devices working correctly and
effectively with entanglement coherence and single-photon. All is
similar to those of the QQTR model (see Section \ref{sec:qqtr_model}).
 
The QUR protocol consists of 5 steps:

\textbf{Step 1:} Preparing, exchanging, and measuring qubits.
\begin{enumerate}
\item For $B_1, .., B_{N+1}$, each prepares $n$ Bell states
  $(|\Phi^+\rangle)^n$.
\item $B_1$ sends the first half of each Bell state to Alice (the
  previous), the second half to $C_1$ (the next). $B_{N+1}$ sends the
  first half of each Bell state to $C_N$ (the previous), the second
  half to Bob (the next). For $i\in[2,N]$, $B_i$ sends the first half
  of each Bell state to $C_{i-1}$ (the previous), the second half to
  $C_i$ (the next).
\item Alice and Bob receive $n$ qubits for each one. They randomly
  choose basis to measure their qubits, independently.
\item Each $C_i, i\in[1,N],$ receives $2n$ qubits from $B_i$ and
  $B_{i+1}$ in a synchronous manner. This means that she receives $n$
  times, and for each time she leads the qubit from $B_i$ and the
  qubit from $B_{i+1}$ to the first and second inputs of the quantum
  circuit as described in Fig. \ref{fig:cnot_m_circuit}. \textsl{Then,
    she sends both two output values to Alice and Bob}. Briefly, $C_i$
  acts exactly as Carol in the QUB protocol.
\item Alice and Bob receive $N$ $2n$-bit strings from $C_1,..,C_N$, and
  informations about positions and basis corresponding.
\item \textsl{The roles of $B_1, ..,B_{N+1}, C_1, .., C_N$ stop here.}
\end{enumerate}

\textbf{Step 2:} Sifting.
\begin{enumerate}
\item Alice and Bob announce their basis.
\item If the basis are different at the position $i$, then Alice,
  Bob discard this position.
\item For each remaining position $i$, Alice and Bob do on $N$
  strings came from $C_1,..,C_N$ as follows. They keep only
  the value of either
  the first output or the second output if their common basis
  is $|\times\rangle$ or $|+\rangle$, respectively.
\item The values of the remaining positions result in $N+2$ $2m$-bit
  strings $a= a_1, ..,a_{2m} ; c(i) = c(i)_1,.., c(i)_{2m}, i=1..N ;
  b=b_1,..,b_{2m}$ where Alice holds $N+1$ string $a, c(1),..,c(N)$ and
  Bob holds $N+1$ string $b, c(1),..,c(N)$. These
  $N+2$ strings should hold $\bigoplus_{j=1}^N c(i)_j = a_i \oplus
  b_i, i\in[1,2m], 2m\sim \frac{n}{2}$.
\end{enumerate}

\textbf{Step 3:} Checking for the presence of Eve.
\begin{enumerate}
\item Alice and  Bob randomly agree $m$ out of $2m$
  positions to check the presence of Eve. This results in two
  $m$-position strings: the check-position string $CP= cp_1,..,cp_{m}$
  and the message-position string $MP= mp_1,..,mp_{m}$.
\item Alice and Bob announce values in check-position $a= a_{cp_1}, ..,a_{cp_m};
  b=b_{cp_1},..,b_{cp_m} ; c(i) = c(i)_{cp_1},.., c(i)_{cp_m}, i=1..N
  $, respectively. They check if $\bigoplus_{i=1}^N c(i)_{cp_j}=a_{cp_j}
  \oplus b_{cp_j}$ or not. If some of negative checks, they abort
  the protocol.
\end{enumerate}

\textbf{Step 4:} Creating the pads for Alice and Bob.
\begin{enumerate}
\item The values in $m$ message-positions result in $N+2$ $m$-bit pads
  $P^A= P^A_1,..,P^A_m$; $P^{C(i)} = P^{C(i)}_1,.., P^{C(i)}_m, i
  \in[1,N]$; and $P^B=P^B_1,..,P^B_m$ where Alice holds $N+1$ pads
  $P^A, P^{C(1)},..,P^{C(N)}$ and Bob holds $N+1$ pads
  $P^B, P^{C(1)},..,P^{C(N)}$. These pads hold $\bigoplus_{i=1}^N P^{C(i)} = P^A \oplus P^B$.
\end{enumerate}

\textbf{Step 5:} Transmitting the key $K$.
\begin{enumerate}
\item Alice creates the random $m$-bit key $K$, $ m\sim
  \frac{n}{4}$. She sends $K \oplus P^A \bigoplus_{i=1}^N P^{C(i)}$ to Bob.
\item Bob receives $K \oplus P^A \oplus \bigoplus_{i=1}^{N}
  P^{C(j)}$, computes $K \oplus P^A \oplus P^B \oplus
  \bigoplus_{i=1}^{N} P^{C(i)} = K$.
\end{enumerate}

\subsubsection{Correctness.}
The QUR protocol is based on the QQTR protocol, therefore, the
correctness is exactly the same.

\subsubsection{Security.}
Note that the random coincidences of basis choices between Alice and
Bob determine the computation basis of EPR states. By the fact that
$C_1,..,C_N$ must announce their measurement values before they know
the basis of Alice and Bob, we have successfully removed the cheating
possibility of $C_1,..,C_N$ as analyzed in the security discussion of
the QUB protocol. Besides, our check process could also
detect imperfect EPR source as that of the modified Lo-Chau BB84
protocol presented in \cite{PSJP00}. In brief, our QUT protocol has the
same security level as the other EPR pair based BB84 protocol.

\section{\label{sec:conc}Conclusion}

We developed quasi-trusted and untrusted models for relaying QKD
keys. We distinguished protocols that are based on single photon and
entangled photons. Our motivation is to avoid difficulties arising
from conserving the quantum entanglement that is unavoidable
dramatically decreased in time. The heart of our works is the quantum
circuit as described as Fig. \ref{fig:cnot_m_circuit}. This circuit
receives two states and gives no more information than the XOR result
of two input states, provided that the two input states are prepared
in a common basis. In particular, the common basis in one out of two
conjugated basis determines the XOR result appearing at either the first
output or the second output. This allows effectively detect malicious
operations on relaying nodes.. 

Our results are very significant. These allow extend the range up to two
time for single photon based QKD and up to un arbitrarily
long distances for entanglement based QKD. Particularly, our
entanglement-based protocols
could keep almost totally the secret-key rate of the
original BB84 protocol. Our protocols do not require having quantum
devices that could keep entangled coherence in long time.

\section{Acknowledgment}
We would like to thank Minh-Dung Dang for helpful discussions.

\bibliographystyle{h-physrev}
\bibliography{qrelay}
\end{document}